\documentclass[epj]{svjour}
\usepackage{graphicx}
\usepackage{amsmath}
\usepackage{amssymb}
\def\slaninafigdir{.}
\begin{document}
\title{%
Dynamical phase transitions in Hegselmann-Krause model of opinion dynamics and consesus
}
\titlerunning{%
Phase transitions in Hegselmann-Krause model
}%
\author{%
Franti\v{s}ek Slanina
\inst{1}%
\thanks{e-mail: {\tt slanina@fzu.cz
}
}}
\institute{
Institute of Physics,
 Academy of Sciences of the Czech Republic,\\
 Na~Slovance~2, CZ-18221~Praha,
Czech Republic\\
and Center for theoretical study, Jilsk\'a 1, Praha, Czech Republic
}
%
%
%
%
%
%
%
\abstract{
The dynamics of the model of agents with limited confidence introduced by
Hegselmann and Krause exhibits multiple well-separated regimes
 characterised by the
number of distinct clusters in the stationary state. We present
indications that there are genuine dynamical phase transitions between
these regimes. The main indicator is the divergence of the average
evolution time required to reach the  stationary state. The slowdown
close to the transition is connected with the emergence of 
the groups of mediator agents which are very small but have decisive role in the
process of social convergence. More detailed study
shows that the histogram of the evolution times is composed of several
peaks. These peaks are unambiguously interpreted as corresponding to
mediator groups consisting of one, two, three etc. agents.
Detailed
study reveals that each transition possesses also an internal fine structure.
}
\maketitle%

\section{Introduction}
Formation of consensus is one of the most studied topics in the field
of sociophysics. It was the subject of the early paper by Callen and
Shapero \cite{cal_sha_74} (which was originally intended as a
contribution to the Moscow seminar banned by the Communist
authorities \cite{chi_ste_75}). The early attempts to apply the ideas
of synergetics to social phenomena were driven by similar ideas
\cite{weidlich_91}. Consensus was in the centre of the papers of Galam
\cite{gal_gef_sha_82,galam_86,gal_mos_91}, who 
revived the term ``sociophysics'' and made it known to general
audience \cite{galam_04c}. For recent reviews, see e. g. 
\cite{cas_for_lor_07,slanina_09a}.

The consensus models can be divided into two well-defined groups. The
models of the first type assume that the agents can choose among a 
small number of discrete 
opinions. The simplest case is the binary choice, studied in the voter
\cite{liggett_99},  
Galam \cite{galam_86,galam_90,galam_99,galam_00}, Sznajd
\cite{szn_szn_00,sta_deo_02,beh_sch_03,sla_lav_03,kru_szn_05,sla_szn_prz_08,lam_red_08},
and majority-rule 
models \cite{kra_red_03,mob_red_03}. 

The second type of models acknowledges that the opinion of the agents
may stretch on a continuous line (or a space of any dimensionality and
structure). The opinions evolve in time by attraction, i. e. the agents
shift their position in the opinion space towards areas where other
agents are already concentrated. Assuming that this dynamics is
linear,  DeGroot  \cite{degroot_74} introduced the model of opinion
convergence in which the opinions in the next time step are linear
combinations of the original opinions. The conditions required for
reaching consensus were clarified in stabilization theorems
\cite{degroot_74,berger_81}. Essentially, the statement is that if the
agents form a network of interactions which is a single connected
cluster, the system always reaches full consensus. The only case in
which different opinions survive in the stationary state is the
trivial one, when the agents split into several clusters with no
communication whatsoever. This is certainly an exaggerated view of the
society as we know it.

\begin{figure}[t]
\includegraphics[scale=0.85]{%
\slaninafigdir/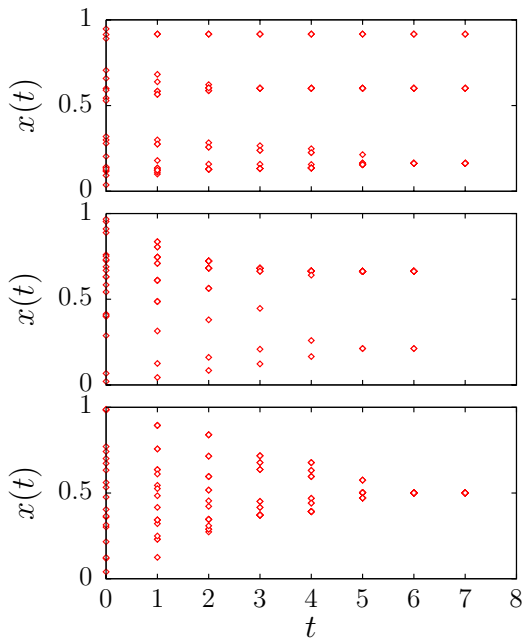}
\caption{
Examples of the evolution of opinions of $N=20$ agents. The confidence
threshold is $\varepsilon=0.1$ (upper panel) and $\varepsilon=0.25$
(middle and lower panel). The evolution is stopped as soon as the
clusters stop changing.
}
\label{fig:evolution-samples-20-1and25}
\end{figure}

The fundamental ingredient missing in the model of DeGroot was the
limited (or bounded) confidence. It is based on a rather trivial
observation that people who differ too much in their opinions are
unable to force the partner shift her opinion and unwilling to make
themselves a
tiniest step towards the opponent. The opinions are frozen, if they are
incompatible. Within discrete-opinion models this idea was excellently
implemented in the Axelrod model 
\cite{axelrod_97,cas_mar_ves_00a,kle_egu_tor_mig_02,vaz_kra_red_02,kle_egu_tor_mig_03a,kle_egu_tor_mig_03b,vaz_red_04,jacobmeier_05,kle_egu_tor_mig_05,gon_cos_tuc_05,kuperman_06,gon_egu_cos_etal_06,vaz_red_06},
while for continuous opinions, bounded confidence was introduced
within the model of Deffuant et
al. \cite{def_nea_amb_wei_00,wei_def_amb_nad_01}. Contrary to the
parallel and linear dynamics of DeGroot, the dynamics in Deffuant et
al. model is stochastic. In each step, a pair of agents is chosen at
random and their opinions are shifted towards each other, on condition
that they do not differ more than the confidence threshold
$\varepsilon$. This model was investigated very thoroughly 
\cite{def_amb_wei_fau_02,benn_kra_red_02,stauffer_02,sta_sou_sch_03,weisbuch_04,sta_mey_04,def_amb_wei_04,amb_def_04,assmann_04,fortunato_04,fortunato_05,wei_def_amb_05}
 both by
simulations of finite systems 
and by numerical solution of the partial integro-differential equation
corresponding to infinite-size limit. It was found that the ultimate
stationary regime is a combination of $\delta$-peaks in the
distribution of opinions. A single peak means full consensus, while
multiple peaks imply breaking the society into several
non-communicating groups. There is a sequence of sharp transitions
between regimes of one, two, three, etc. peaks, at critical values of
the confidence threshold. Numerical estimates suggest that the
transition from full consensus to multiple peaks occurs at
$\varepsilon_{c1}\simeq 0.5$. However, the side peaks only gain
macroscopic weight at another critical value $\varepsilon_{c2}\simeq 0.27$
\cite{benn_kra_red_02,fortunato_04,lorenz_07}.

While the model of Deffuant et al. uses sequential stochastic
dynamics, the model of Hegselmann and Krause (HK) \cite{heg_kra_02} 
is more close to the
original DeGroot model. The randomness enters only in the initial
condition and further evolution is deterministic. In each step, the new
values of the opinion variable are linear combinations of those
opinions, which are not farther than the confidence threshold. 
From the uniformly random initial condition, one or several groups of
identical opinions evolve. Contrary to the Deffuant et al. model, the
absorbing state (i. e. such that none of the opinions can change any
more) is reached after finite number of steps, provided the number of
agents is finite. The HK model was investigated both by simulations
and by solution of corresponding partial integro-differential equation 
\cite{lorenz_07,plu_lat_rap_05,fortunato_04b,fortunato_04a,fortunato_05a,fortunato_05b,for_sta_05,for_lat_plu_rap_05,heg_kra_06,lorenz_07a,lorenz_08,lorenz_08a}.
Numerically, it was found that the transition to
full consensus appears around the critical value $\varepsilon_c\simeq
0.2$ \cite{fortunato_05a}. A smart way of discretization the
integro-differential equation, called interactive Markov chain 
\cite{lorenz_07a,lorenz_08,lorenz_08a,lorenz_06,lorenz_07b},
provides two conflicting results for the consensus transition. For odd
number of discretization intervals, the answer is $\varepsilon_c\simeq
0.19$  \cite{lorenz_07,lorenz_07a}, while for even number of intervals one gets $\varepsilon_c\simeq
0.22$ \cite{for_lat_plu_rap_05,lorenz_07a}. Later, we shall mention arguments 
indicating that the correct discretization is with odd number of intervals.
The advantage of the approach using interactive Markov chains is that
in enables proving stabilization theorems on the HK dynamics
\cite{lorenz_05,lorenz_07c,lor_lor_08}.

Various modifications of  Deffuant et al. and HK models
 were investigated. For example, a model
which interpolates between Deffuant et al. and HK was introduced
\cite{urb_lor_07}. Heterogeneous confidence thresholds
\cite{lorenz_08,lag_abr_zan_04},
 influence of extremists \cite{def_amb_wei_fau_02,por_bol_sti_07}
and presence of a ``true truth''  \cite{heg_kra_06,malarz_06} were
 studied. Introduction of
multi-dimensional opinion space
\cite{wei_def_amb_nad_01,for_lat_plu_rap_05,lorenz_06,lag_abr_zan_03}
is also a  natural generalization. Interestingly, introduction of
noise into the dynamics alters the behaviour profoundly
\cite{pin_tor_her_09}. This might be interpreted so that HK and
Deffuant et al. models follow a strictly zero-temperature dynamics,
which is unstable with respect to noise.

The aim of this paper is to investigate in detail the transitions from full
consensus to state with two groups, to state with three, four
etc. groups. Especially, we show in detail the phenomenon of critical
slowdown, already hinted in
\cite{for_lat_plu_rap_05,lorenz_06,urb_lor_07} and show how it is
related to the presence of mediators, introduced on an intuitive level
in \cite{lorenz_07b}.

\begin{figure}[t]
\includegraphics[scale=0.85]{%
\slaninafigdir/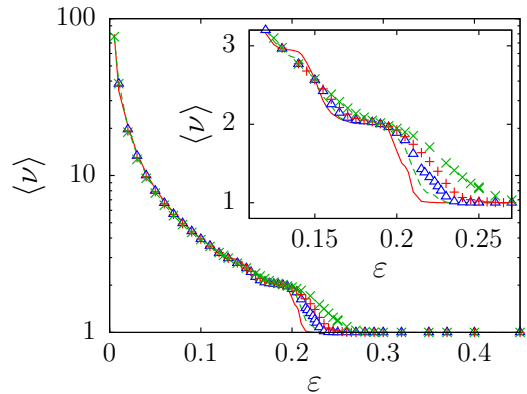}
\caption{
Dependence of the average number of clusters on the confidence
threshold. The number of agents is $N=5000$ (solid line), $2000$
(dashed line), $1000$ ($\bigtriangleup$), $500$ ($+$), and $200$
($\times$). In the inset, detail of the same data.}
\label{fig:nuclusters-all}
\end{figure}

\begin{figure}[t]
\includegraphics[scale=0.85]{%
\slaninafigdir/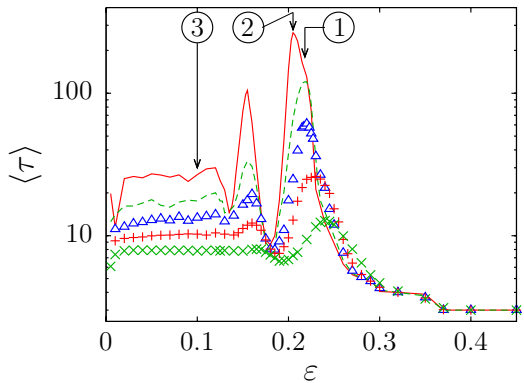}
\caption{
Dependence of the average time to reach an absorbing state 
on the confidence
threshold. The number of agents is $N=5000$ (solid line), $2000$
(dashed line), $1000$ ($\bigtriangleup$), $500$ ($+$), and $200$
($\times$). The arrows with circled numbers indicate the values of
$\varepsilon$ used in Figs. \ref{fig:constime-histogram-2000-218},
\ref{fig:constime-histogram-5000-205}, and \ref{fig:constime-histogram-all-1}.}
\label{fig:constime-all}
\end{figure}

\section{Phases in the Hegselmann-Krause model}

\subsection{Definitions}

Let us first recall the definition of the HK model. The system
consists of $N$ agents. The opinion of agent $i$ at time $t$ is a number $x_i(t)\in
(0,1)$.  Thus, the state of the system is described by the
$N$-component vector $x(t)$. The evolution of the state vector in
discrete time $t=0,1,2\ldots$
is
deterministic and seemingly linear
\begin{equation}
x_i(t+1)=\sum_{j=1}^N M_{ij}[x(t)]\,x_j(t)
\label{eq:hk-definition}
\end{equation}
but the mixing matrix $M$ is not constant, but depends on the actual state
$x$. The dependence $M[x]$ is dictated by the principle of bounded
confidence. If $\varepsilon\in(0,1)$ is the confidence threshold, then 
\begin{equation}
M_{ij}[x]=
\left\{
\begin{array}{ll}
0&\mathrm{ for}\;\;|x_i-x_j|>\varepsilon\\
\frac{1}{N_{ij}}&\mathrm{ for}\;\;|x_i-x_j|\le\varepsilon
\end{array}
\right.
\label{eq:mixing-matrix-definition}
\end{equation}
where the normalization factor $N_{ij}$ is the number of agents not
farther than $\varepsilon$ from the agent $i$,
$N_{ij}=|\{j:|x_i-x_j|\le\varepsilon\}|$. 
As the initial condition, we choose set of independent random values
$x_i(0)$, uniformly distributed in the interval $(0,1)$.

The dynamics (\ref{eq:hk-definition}), (\ref{eq:mixing-matrix-definition})
 has infinite number of absorbing
states. They can be classified according to the number of
non-communicating clusters. The state with $\nu$ clusters is
characterised by numbers $f_1<f_2<\ldots<f_{\nu}$ such that
$f_{l+1}-f_l>\varepsilon$ and $\forall i\,\exists l:x_i=f_l$. The
smallest $t$ for which $x(t)$ is an absorbing state will be called
consensus time and denoted $\tau$.

As the initial condition is random, the time  $\tau$ to reach an
absorbing state as well as the number $\nu$ of clusters in that state
are also random variables. We shall be mainly interested in the mean
values $\langle\tau\rangle$ and $\langle\nu\rangle$, averaged over
initial conditions.

\begin{figure}[t]
\includegraphics[scale=0.85]{%
\slaninafigdir/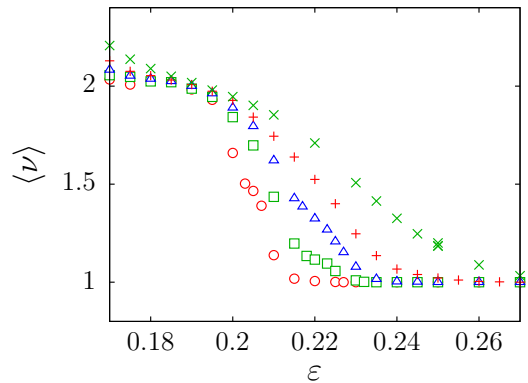}
\caption{
Detail of the dependence of the average number of clusters on the confidence
threshold. The number of agents is $N=5000$ ({\Large $\circ$}), $2000$
($\Box$), $1000$ ($\bigtriangleup$), $500$ ($+$), and $200$
($\times$). }
\label{fig:nuclusters-all-detail}
\end{figure}

\begin{figure}[t]
\includegraphics[scale=0.85]{%
\slaninafigdir/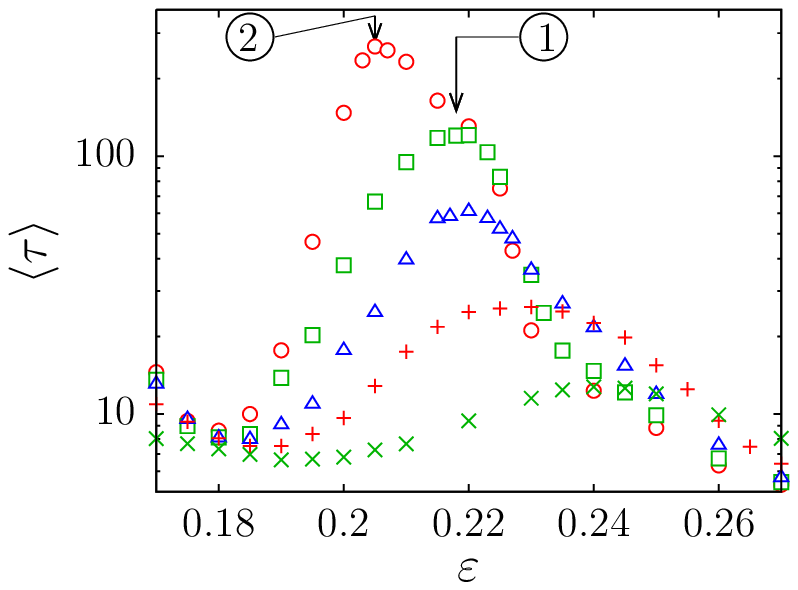}
\caption{
Detail of the dependence of the average  time to reach an absorbing
state 
on the confidence
threshold. The number of agents is $N=5000$ ({\Large $\circ$}), $2000$
($\Box$), $1000$ ($\bigtriangleup$), $500$ ($+$), and $200$
($\times$). The arrows with circled numbers indicate the values of
$\varepsilon$ used in Figs. \ref{fig:constime-histogram-2000-218} and
\ref{fig:constime-histogram-5000-205}.}
\label{fig:constime-all-detail}
\end{figure}

\subsection{Which absorbing state?}

The number of clusters in the absorbing state depends mainly on the
confidence threshold $\varepsilon$, but also on the initial
condition. We show in Fig. \ref{fig:evolution-samples-20-1and25} three
typical examples. For large enough $\epsilon$ the evolution ends in a
state with single cluster, while for smaller $\varepsilon$ the
resulting $\nu$ differs according to the configuration of opinions at
the beginning. If we average the final number of clusters, we observe
a decreasing function of $\varepsilon$, as shown in
Fig. \ref{fig:nuclusters-all}. A more detailed look (see the inset in
Fig. \ref{fig:nuclusters-all}) shows that for increasing number of agents,
 well-defined plateaus
develop at integer values of $\langle\nu\rangle$, separated by steps
which become sharper for increasing $N$ and we may conjecture that
discontinuities emerge for $N\to\infty$ at critical values
$\varepsilon=\varepsilon_{c1}$, $\varepsilon_{c2}$, etc. From Fig.
\ref{fig:nuclusters-all}) we can estimate the first two of them as
$\varepsilon_{c1}\simeq 0.2$,  $\varepsilon_{c2}\simeq 0.14$.

\subsection{Critical slowing down}

The critical values $\varepsilon_{ck}$ mark dynamical phase
transitions from regime with $k$ clusters in absorbing state to $k+1$
clusters. It is very questionable if the notions of first-order versus
continuous phase transitions can be transferred from equilibrium to
non-equilibrium transitions. However, we can study certain features,
which are distinctive in equilibrium, also in non-equilibrium
case. One of them is  the slowdown of the dynamics close to the
critical point. This is a signature of continuous transition. In HK
model, we can measure the average time to reach an absorbing state as a function of
$\varepsilon$, and indeed, we observe peaks located at the transition
regions, as seen in Fig. \ref{fig:constime-all}. The height of the
peaks increases 
with the number of agents, which suggests diverging  time at
the transition points. The overall picture emerging from these results
seems to be the following. In HK model in the limit $N\to\infty$,
 we have a sequence of phases
characterised by one, two, three. etc. clusters in the absorbing state
which is the result of the dynamics. The phase transitions occur at
confidence thresholds $\varepsilon_{c1}$,  $\varepsilon_{c2}$ etc,
where the average number of clusters jumps discontinuously between two
integer values, and where the average consensus time diverges. Having
this in mind, we can consider the phase transitions second-order. In
the following sections we shall see that the phase transitions in HK
model are even more subtle than that.

\begin{figure}[t]
\includegraphics[scale=0.85]{%
\slaninafigdir/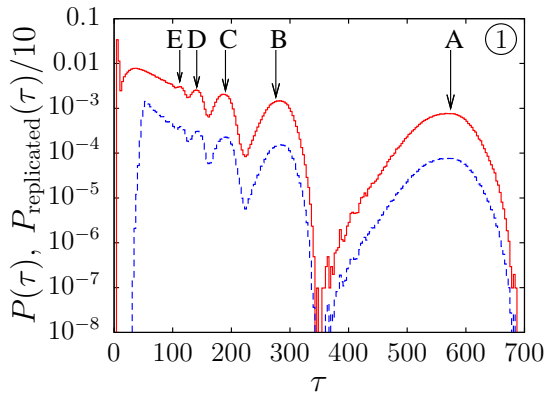}
\caption{
Histogram of  times to reach an absorbing state, for $N=2000$ and
$\varepsilon=0.218$ (full line). The arrows marked by capital letters A to E
indicate the length of consensus time realised in the evolution
samples shown in Fig. \ref{fig:evolution-samples-2000-218}. The
circled ``$1$'' refers to the arrow in Figs. \ref{fig:constime-all}
and \ref{fig:constime-all-detail}. We draw also the
distribution found by replication of the longest peak, according to
(\ref{eq:replicated-peak}), with $k_\mathrm{max}=11$ (dashed line).
 For better visibility, it is scaled down by the factor $10$.} 
\label{fig:constime-histogram-2000-218}
\end{figure}

\begin{figure}[t]
\includegraphics[scale=0.85]{%
\slaninafigdir/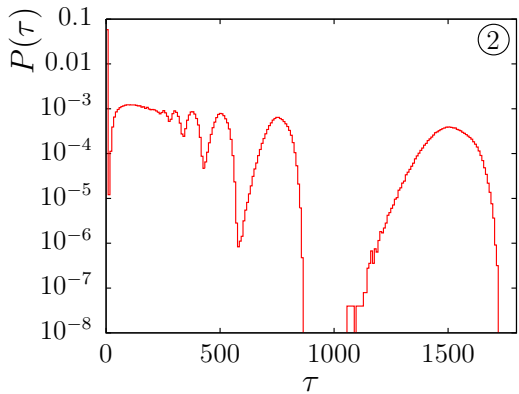}
\caption{
Histogram of  times to reach an absorbing state, for $N=5000$ and
$\varepsilon=0.205$. }
\label{fig:constime-histogram-5000-205}
\end{figure}

\begin{figure}[t]
\includegraphics[scale=0.85]{%
\slaninafigdir/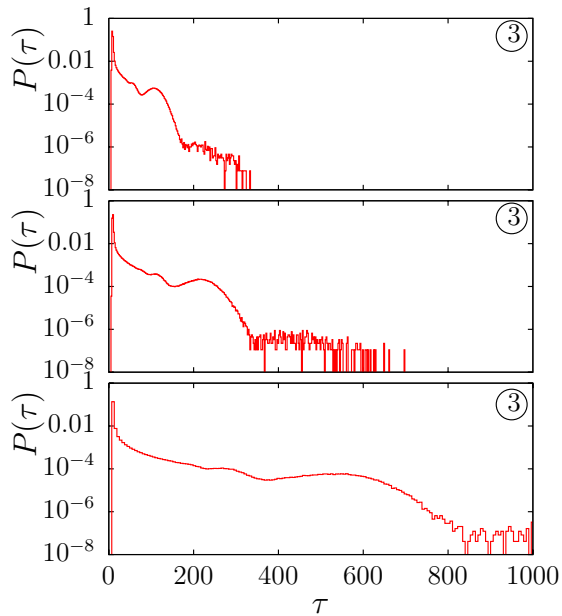}
\caption{
Histogram of times to reach an absorbing state, for 
$\varepsilon=0.1$ and $N=1000$ (upper panel), $N=2000$ (middle
panel), and $N=5000$ (lower panel).}
\label{fig:constime-histogram-all-1}
\end{figure}

\section{How the absorbing state is reached}

From now on, we shall concentrate on the first of the sequence of
transitions, where the full consensus ends. We show in Figs
\ref{fig:nuclusters-all-detail}  and
\ref{fig:constime-all-detail} 
 details of the $\varepsilon$-dependence of  average number of
clusters and average time to reach an absorbing state, respectively. 

We can see in Fig. \ref{fig:nuclusters-all-detail} that increasing $N$
results in decrease of $\langle\nu\rangle$ in the transition
region. (We shall defer the sociological perspective of this phenomenon
to the Conclusions.) The transition becomes steeper, but the inflexion
point is shifted leftwards. Similarly, in
Fig. \ref{fig:constime-all-detail} we observe that the peak not only
grows when number of agents increases, but shifts quite markedly to
lower values of $\varepsilon$. The vales of $\varepsilon_{c1}$
inferred from the finite-$N$ results must be considered as upper
bounds to the true critical value valid in the thermodynamic limit.

We can gain further insight into the divergence of consensus time at
the transition, if we plot the histogram of times to reach
an absorbing state for values
of $\varepsilon$ close to the maximum of the peak in
$\langle\tau\rangle$. We show the results for $N=2000$ at
$\varepsilon=0.218$ and for $N=5000$ at $\varepsilon=2.05$, in
Figs. \ref{fig:constime-histogram-2000-218} and 
\ref{fig:constime-histogram-5000-205}, respectively. The
characteristic feature of the histograms
is a sequence of peaks. The height of the peaks is
nearly the same, especially for larger $N$. For comparison, we plot in
Fig. \ref{fig:constime-histogram-all-1} the histogram of consensus
times for $\varepsilon=1$, far from any major peak in
$\langle\tau\rangle$. There are barely visible traces of peaks, but as
the system size increases, the histogram becomes flat, contrary to the
transition region, where the peaks in the histogram become more
pronounced. Therefore, the peaks in the histogram are tightly related
to the divergence of consensus time at the transition. 

\begin{figure}[t]
\includegraphics[scale=0.85]{%
\slaninafigdir/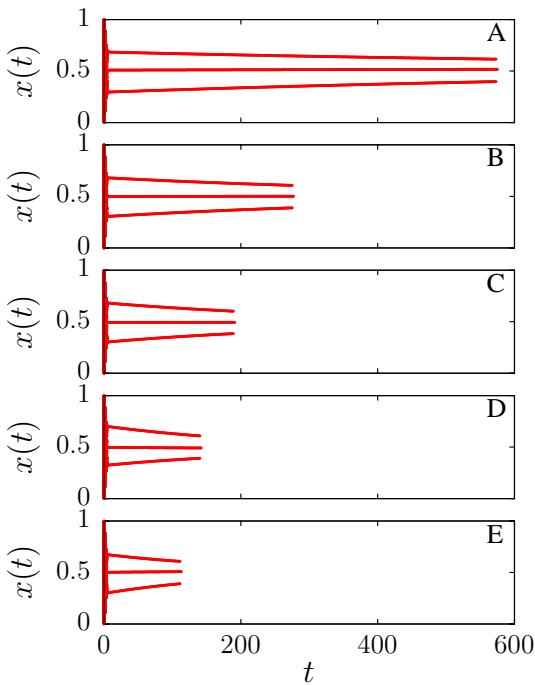}
\caption{
Examples of the evolution with different size of the central mediator
group, from top to bottom $N_\mathrm{med}=1$, $2$, $3$, $4$, and
$5$. The capital letters in the top right corners relate to the arrows
in Fig. \ref{fig:constime-histogram-2000-218}.}
\label{fig:evolution-samples-2000-218}
\end{figure}

As a next step, we must ask what is the origin of the peaks. The
emergence of the peaks implies that there are certain typical lengths
of the evolution from the initial condition to the absorbing state. We
naturally expect that the typical lengths correspond to typical
structural features of the evolution. To see that, we show in
Fig. \ref{fig:evolution-samples-2000-218} spatio-temporal diagrams
of the evolution of the system for five principal peaks in the
histogram. The consensus times are indicated by letters A to E in
Fig.  \ref{fig:constime-histogram-2000-218} and the
corresponding panels in
Fig.  \ref{fig:evolution-samples-2000-218} are denoted by the same
letters. We can see immediately a 
 common feature of
all these five samples. After a very short transient period, three
clusters are formed, one of them close to the exact middle and two of
them on the wings. The latter are slowly attracted to the central
cluster, until their distance falls below $\varepsilon$. Then, all three
collapse into a single cluster and an absorbing state with full
consensus is reached.

Neglecting the very short transient, the consensus time is given by
the time needed to attract the wing clusters to the distance
$\varepsilon$. We assume that the middle cluster contains
$N_\mathrm{med}$ ``mediator'' agents and is located at $x(0)=1/2$, while
the other groups are equal in size $N_+=N_-=(N-N_\mathrm{med})/2$ and
are located initially at $x_\pm(0)=1/2\pm \Delta x$. The middle cluster does
not move and  the wing clusters evolve according to the difference equation
\begin{equation}
x_\pm(t+1)-x_\pm(t)=-\frac{N_\mathrm{med}}{N_\pm+N_\mathrm{med}}\bigg(x_\pm
-\frac{1}{2}\bigg)
\;.
\label{eq:three-clusters-evolution-discrete}
\end{equation}
For $N_\mathrm{med}/N\ll 1$ the dynamics is very slow and we can
replace the difference in (\ref{eq:three-clusters-evolution-discrete})
by derivative. Hence, the consensus time is estimated as
\begin{equation}
\tau=\frac{N}{2N_\mathrm{med}}\,\ln\frac{2\Delta x}{\varepsilon}
\;.
\label{eq:three-clusters-evolution-constime}
\end{equation}
Since the initial condition must be $\Delta x <\varepsilon$, otherwise
the clusters would never coalesce, and $N_\mathrm{med}\ge 1$, we get a
strict upper bound to the consensus time, provided  the mechanism of
three clusters is in force
\begin{equation}
\tau \le N\,\ln\sqrt{2}
\;.
\label{eq:three-clusters-evolution-constime-bound}
\end{equation}
Indeed, we can see that the histograms in Figs.  \ref{fig:constime-histogram-2000-218} and 
\ref{fig:constime-histogram-5000-205} obey the 
bound (\ref{eq:three-clusters-evolution-constime-bound}).

The width of the peaks in the histogram is due to the fluctuations in
the initial positions of the wing clusters. The peaks differ only
in the number of mediators. Indeed, the evolution patterns A to E in
Fig.  \ref{fig:evolution-samples-2000-218} are observed for number of
mediators $1,2,\ldots,5$. Comparing that with Fig.
\ref{fig:constime-histogram-2000-218}, where the peaks are denoted by
corresponding letters A to E, we clearly see that the peak at longest
consensus has $N_\mathrm{med}=1$, the second has $N_\mathrm{med}=2$
etc. This fact suggests, that the peaks for
$N_\mathrm{med}=2,3,\ldots$ can be obtained by replication the peak at
$N_\mathrm{med}=1$. Denoting $P_1(\tau)$ the latter peak only, we
approximate the full distribution of consensus times by
\begin{equation}
P(\tau)\simeq P_\mathrm{replicated}(\tau)=
\sum_{k=1}^{k_\mathrm{max}} k\,P_1(k\,\tau)\;.
\label{eq:replicated-peak}
\end{equation}
This approximation assumes that all sizes of the mediator group up to
$N_\mathrm{med}=k_\mathrm{max}$ have the same probability and neglect
the influence of the initial short transient. Therefore, it is
reasonably accurate for a few highest peaks, but fails at short
$\tau$, as it is confirmed in Fig.  \ref{fig:constime-histogram-2000-218}. 

Let us also note that the mechanism of mediators located in the middle
explains why, in the numerical solution of the partial differential equation for
HK model, the discretization into even number of equally-sized
intervals is wrong. Indeed, in this case the mediator cluster is
located just at the border of two intervals, however fine the
discretization is, and this induces numerical artifacts into the results.

\section{Fine structure of the transitions}

We already noted that the dependence of $\langle\nu\rangle$ on
$\varepsilon$ is not like the dependence of average magnetization on
temperature, as seen in simulations of finite-size Ising model. The
transition region is not only squeezed into more narrow region, but is
also shifted to lower $\varepsilon$. The same is observed also in
$\langle\tau\rangle$. When the system size grows, the peaks do not
simply grow and get thinner, but are also shifted to  lower
$\varepsilon$, consistently with the behaviour of
$\langle\nu\rangle$. Let us look at this shifting of peaks in more
detail. 

\begin{figure}[t]
\includegraphics[scale=0.85]{%
\slaninafigdir/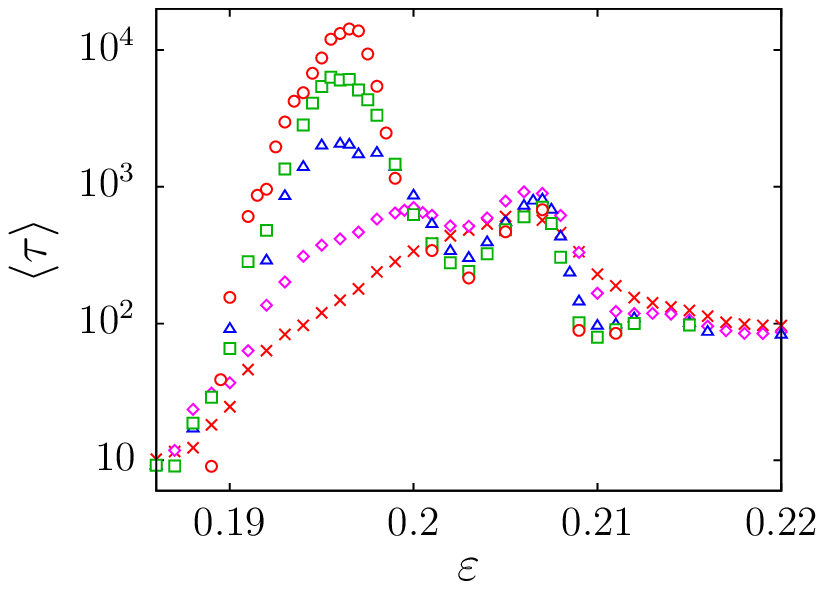}
\caption{
Fine structure of the average time to reach an absorbing state, at the 
transition from full consensus to phase with two
clusters. The system size is $N=2\cdot 10^5$  ({\Large $\circ$}),
$10^5$ ($\Box$), $ 5\cdot 10^4$
($\bigtriangleup$),  $2\cdot 10^4$ ($\diamond$),
$10^4$  ($\times$).   
}
\label{fig:constime-finestructure}
\end{figure}

\begin{figure}[t]
\includegraphics[scale=0.85]{%
\slaninafigdir/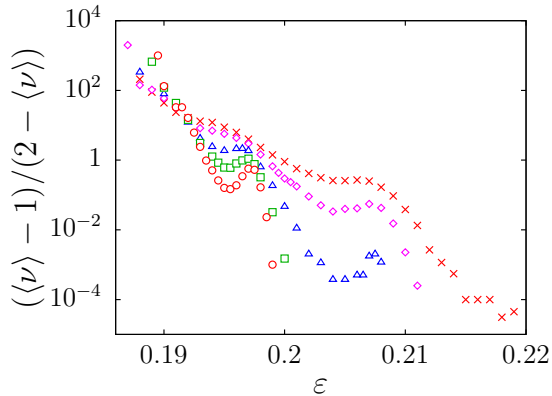}
\caption{
Fine structure of the average number of clusters in absorbing state, at the 
transition from full consensus to phase with two
clusters. The system size is $N=2\cdot 10^5$  ({\Large $\circ$}),
$10^5$ ($\Box$), $ 5\cdot 10^4$
($\bigtriangleup$),  $2\cdot 10^4$ ($\diamond$),
$10^4$  ($\times$).   
}
\label{fig:nuclusters-finestructure}
\end{figure}

To this end, we performed simulations of fairly  large systems (up to
$N=2\cdot 10^5$) in the range of $\varepsilon$ which covers the
transition from the full consensus phase ($\langle\nu\rangle=1$) to
the phase with two clusters  ($\langle\nu\rangle=2$). The picture
which emerges, is demonstrated in
Figs. \ref{fig:constime-finestructure} and
\ref{fig:nuclusters-finestructure}. It is somewhat surprising that the
peaks in $\langle\tau\rangle$ only apparently move. Closer look at
Fig. \ref{fig:constime-finestructure}  reveals that a peak at certain
value of $\varepsilon$ remains at the same position when $N$ grows,
but a new peak starts growing at somewhat smaller $\varepsilon$. When
this second peak reaches some height, it saturates and another peak is
born and grows at even smaller $\varepsilon$. In this way, older peaks
do not depend on $N$ any more, but rather
are overgrown by new ones. To our knowledge, this effect has no
analogy in equilibrium phase transitions and is entirely related to
dynamical nature of the transition in HK model.

Similar fine structure of the transition region is observed on the
dependence of average number of clusters on $\varepsilon$. In the
transition region, it drops from $\langle\nu\rangle=2$ 
to  $\langle\nu\rangle=1$. To make
the details more visible, we plot the
quantity $(\langle\nu\rangle-1)/(2-\langle\nu\rangle)$,  instead of
$\langle\nu\rangle$, in logarithmic
scale. In Fig. \ref{fig:nuclusters-finestructure} we can see that
$\langle\nu\rangle$ drops from $2$ to $1$ in step-wise manner. For
$N=10^4$ we observe plateaus, or regions of $\varepsilon$, where the
average number of clusters is nearly constant somewhere between $1$
and $2$. When the system size grows, these steps, or plateaus,
diminish in the value of $\langle\nu\rangle$ but keep their
width. Moreover, the edges of the steps decrease more slowly, so that
the dependence of $\langle\nu\rangle$ on $\varepsilon$ becomes
non-monotonous and the ``plateaus'' have depression in the middle.
 Interestingly, the peaks in $\langle\tau\rangle$ are located just
 next to the right edges of these ``plateaus''. We assume that the
 sequence of the peaks in $\langle\tau\rangle$  and plateaus in
 $\langle\nu\rangle$ tends to a point
 $\varepsilon=\varepsilon_{c1}$,
 which is the location of the true phase transition in the
 limit $N\to\infty$.  Form the data in Figs. 
  \ref{fig:constime-finestructure} and
\ref{fig:nuclusters-finestructure} 
we can estimate $\varepsilon_{c1}\simeq 0.19$.

Comparing Figs.  \ref{fig:constime-finestructure} and
\ref{fig:nuclusters-finestructure} we can see that the non-monotonous
dependence of $\langle\nu\rangle$ on $\varepsilon$ goes hand in hand
with the multiple-peak dependence of  $\langle\tau\rangle$ on
$\varepsilon$. We do not have a detailed account for this phenomenon,
but the following scenario seems plausible.

The behaviour in the transition region is dominated by the slow
evolution of three-cluster system, as described above. The existence
of full consensus depends on emergence of the mediator group. In other
words, the average number of clusters is related to the probability
$P_\mathrm{med}(\varepsilon;0)$ 
that the mediator group is empty, as
$\langle\nu\rangle=1+P_\mathrm{med}(\varepsilon;0)$, as long as 
more than two clusters  in the absorbing state occur with negligible
probability.
We suppose that for given $\varepsilon$ and very large $N$ the
fraction
 of agents in the mediator group approaches a limit
$\mu(\varepsilon)=\lim_{N\to\infty} N_\mathrm{med}/N$. As the full
 consensus is only possible if $\mu(\varepsilon)>0$, we may consider
 $\mu(\varepsilon)$ as order parameter of the non-equilibrium phase
 transition in HK model. The location $\varepsilon_{c1}$ of the transition 
is determined by $\mu(\varepsilon_{c1})=0$.

We also assume that a
 ``master'' probability distribution exists $F(\rho,n)$, independent of
 $\varepsilon$ and $N$, so that the probability distribution for
 $N_\mathrm{med}$ is
\begin{equation}
P_\mathrm{med}(\varepsilon;N_\mathrm{med})= F(\mu(\varepsilon)\,N;N_\mathrm{med})\;.
\end{equation}
The parameter $\rho$ stands for the average size of the mediator
group, so $\rho=\sum_{n=1}^\infty n F(\rho,n)$.  We do not have direct
access to the distribution $F(\rho,n)$ in 
simulations. In absence of any other information we hay hypothesise
that the distribution might be Poissonian,
$F(\rho,n)=\mathrm{e}^{-\rho}\,\rho^n/n!$. According to
(\ref{eq:three-clusters-evolution-constime}) and assuming that $\Delta
x$ is proportional to $\varepsilon$, we have the estimate
\begin{equation}
\langle\tau\rangle\propto 
\sum_{N_\mathrm{med}=1}^{N_\mathrm{med,max}}\frac{N}{N_\mathrm{med}}
F(\mu(\varepsilon)\,N;N_\mathrm{med}) \;.
\label{eq:consensustime-analytical-estimate}
\end{equation}
The upper bound $N_\mathrm{med,max}$ for the size of the mediator
group can be safely extended to infinity. For fixed $\varepsilon$ (and therefore
fixed $\mu(\varepsilon)$) and $N\to\infty$ the average consensus time
approaches a limit which is proportional to $\langle\tau\rangle\propto
1/\mu(\varepsilon)$. On the other hand, for $N$ fixed and variable
$\varepsilon$, the dependence of $\langle\tau\rangle$ according
to (\ref{eq:consensustime-analytical-estimate}) develops a maximum as
a function of $\mu$. The location of the maximum shifts when $N$ grows
as $\mu_\mathrm{max}\propto 1/N$. This way, the location of the peak in
$\langle\tau\rangle(\varepsilon)$ approaches $\varepsilon_{c1}$ as
$N\to\infty$. 

If the fraction $\mu$ of agents in
the mediator cluster was a mono\-tonous function of $\varepsilon$, with
$\mu=0$ at the critical point $\varepsilon=\varepsilon_{c1}$, we would
see a peak in $\langle\tau\rangle$ growing and shifting
gradually to lower values of $\varepsilon$, up to its asymptotic
position at the critical point. Then, also $\langle\nu\rangle=1+F(\mu(\varepsilon)\,N;0)$ would
be a monotonously decreasing function of $\varepsilon$. However, we
can see violation of this monotonicity in
Fig. \ref{fig:nuclusters-finestructure}. Therefore, $\mu$ is not 
a monotonous function of $\varepsilon$, which explains both the
non-monotonicity of $\langle\nu\rangle$ and the fact that multiple
peaks appear in $\langle\tau\rangle$, instead of observing smooth
shift and growth of a single peak. The non-monotonicity imposes a
deformation on the otherwise smooth growth and shift of the peak in
$\langle\tau\rangle$. This deformation results in apparent emergence
of new peaks next to the older ones. In fact, as long as the
approximation $\lim_{N\to\infty}\langle\tau\rangle\propto
1/\mu(\varepsilon)$ 
is justified, the non-monotonicity in $\mu(\varepsilon)$ is directly
visible in non-monotonicity, i. e. multiple-peak structure, 
 of $\langle\tau\rangle$, close to the critical point. 

 However, the key ingredient of the
whole phenomenon of fine structure of the transition, which is the
non-monotonicity of $\mu(\varepsilon)$ remains unexplained. Clearly,
it relies on the processes happening within the relatively short
transient period. The three-cluster
structure, i. e. two wings plus mediators, is formed in this period
and and the distribution of the number of
mediators is established, which we assumed, for simplicity, to have the form
$F(\mu(\varepsilon)\,N;N_\mathrm{med})$, but actually can be more
complex.

\section{Conclusions}

We investigated in detail phase structure of the Hegselmann-Krause model
of consensus formation. The only parameters of the model are
confidence threshold and number of agents. The dynamics is
deterministic, but the initial condition is random.
 We found that, depending on the value of the confidence threshold,
 well-defined phases exist,
characterised by the number of non-communicating clusters in the
absorbing state. This number is 
 one in full consensus phase, while it is two, three, etc. in phases
 lacking full consensus among all agents, but exhibiting
 consensus within the clusters. The phases are separated
by dynamical phase transitions, characterised by divergence of the
 time needed to reach the absorbing state, reminiscent of critical
 slowing down known from 
second order equilibrium phase transitions. 

The mechanism which leads to the divergence of characteristic time at
the phase transition is related to the emergence of a group of
mediators, i. e. a small cluster in the middle of the opinions, which
is able to attract the two clusters on the left and right wings from the
mediators. The mediator cluster can be arbitrarily small, but
non-empty. One single mediator is able to attract arbitrarily large
wing clusters, if they are located initially within the confidence threshold.
The attraction is the slower the larger the wing clusters are, but
typically close to the transition the wing clusters contain nearly all
the agents, while the fraction contained in the mediator cluster is
tiny. Hence
the divergence of the time needed to reach the absorbing state, when
the system size grows. This mechanism is reflected also in the
histogram of times to reach consensus, which exhibits a characteristic
series of peaks. Each of the peaks corresponds to a specific number of
agents in the mediator group, which is one for the farthest peak, two
for the next one, etc.

The most surprising feature of the dynamical phase transition in HK
model is its fine structure. In the transition region, the average
time to reach absorbing state, as a function of the confidence
threshold,
 exhibits not just a growing peak when system size grows. The
peak is also shifted towards lower values, in a complex
manner. Apparently, the peak grows with system size until saturation,
and then a new peak starts growing at a lower value of the confidence
threshold. Thus, a series of peaks, overgrowing each other,
emerges. We assume that the positions of the peaks tend to a limit
which is the location of the phase transition in the infinite-size limit.

If we interpret the results obtained in terms of the (hypothetical)
average fraction of agents in the mediator cluster, we come to
conclusion that this quantity must be a non-monotonous function of the
confidence threshold in the transition region. If it were monotonous,
the peak in the average time to reach absorbing state would
continuously shift towards lower values when system size grows. But
non-monotonicity of the average size of the mediator cluster imposes a
deformation on this shift, which looks like new peaks were born next
to older ones. However, we must admit that the non-monotonicity of the
average fraction of agents in the mediator cluster remains unexplained.

Finally, let us make one sociological observation. In the transition
region from full consensus phase, the average number of clusters in
the absorbing state reflects the probability to reach consensus. When
the system size grows, with confidence threshold fixed, the
probability of consensus increases. More agents are more likely to
reach consensus at the end. It is easy to understand this phenomenon
in terms of the mediators. In a larger system of agents the
probability to get non-empty mediator group is larger. Because this
tiny mediator group is vital for consensus, it is easier to reach
consensus in larger society. It is a challenge to experimental
sociologists to test this prediction in reality.
\begin{acknowledgement}
This work was carried out within the project AV0Z10100520 of the Academy 
of Sciences of the Czech republic and was  
supported by the M\v SMT of the Czech Republic, grant no. 
OC09078 and by the Research Program CTS MSM 0021620845.

\end{acknowledgement}

\begin{thebibliography}{99}
\bibitem{cal_sha_74}
E. Callen and D. Shapero,
Phys. Today
 23
 (July 1974).

\bibitem{chi_ste_75}
N. A. Chigier and E. A. Stern (Editors),
{\it Collective phenomena and the applicatios of physics to other fields of science}
 (Brain Research Publications, Fayetteville, 1975).

\bibitem{weidlich_91}
W. Weidlich,
Physics Reports
 {\bf 204},
 1
 (1991).

\bibitem{gal_gef_sha_82}
S. Galam, Y. Gefen, and Y. Shapir,
J. Math. Sociol.
 {\bf 9},
 1
 (1982).

\bibitem{galam_86}
S. Galam,
J. Math. Psychol.
 {\bf 30},
 426
 (1986).

\bibitem{gal_mos_91}
S. Galam and S. Moscovici,
Eur. J. Soc. Psychol.
 {\bf 21},
 49
 (1991).

\bibitem{galam_04c}
S. Galam,
Physica A
 {\bf 336},
 49
 (2004).

\bibitem{cas_for_lor_07}
C. Castellano, S. Fortunato, and V. Loreto,
Rev. Mod. Phys.
 {\bf 81},
 591
 (2009).

\bibitem{slanina_09a}
F.\ Slanina,
in: {\it  Encyclopedia of Complexity and Systems Science},
 8379
 (Springer, New York, 2009).

\bibitem{liggett_99}
T. M. Liggett,
{\it Stochastic Interacting Systems: Contact, Voter, and Exclusion Processes}
 (Springer, Berlin, 1999).

\bibitem{galam_90}
S. Galam,
J. Stat. Phys.
 {\bf 61},
 943
 (1990).

\bibitem{galam_99}
S. Galam,
Physica A
 {\bf 274},
 132
 (1999).

\bibitem{galam_00}
S. Galam,
Physica A
 {\bf 285},
 66
 (2000).

\bibitem{szn_szn_00}
K. Sznajd-Weron and J. Sznajd,
Int. J. Mod. Phys. C
 {\bf 11},
 1157
 (2000).

\bibitem{sta_deo_02}
D. Stauffer and P. M. C. de Oliveira,
cond-mat/0208296.

\bibitem{beh_sch_03}
L. Behera and F. Schweitzer,
Int. J. Mod. Phys. C
 {\bf 14},
 1331
 (2003).

\bibitem{sla_lav_03}
F. Slanina and H. Lavi\v{c}ka,
Eur. Phys. J. B
 {\bf 35},
 279
 (2003).

\bibitem{kru_szn_05}
S. Krupa and K. Sznajd-Weron,
Int. J. Mod. Phys. C
 {\bf 16},
 177
 (2005).

\bibitem{sla_szn_prz_08}
F.\ Slanina, K.\ Sznajd-Weron and P.\ Przyby{\l}a,
Europhys. Lett.
 {\bf 82},
 18006
 (2008).

\bibitem{lam_red_08}
R. Lambiotte and S. Redner,
Europhys. Lett.
 {\bf 82},
 18007
 (2008).

\bibitem{kra_red_03}
P. L. Krapivsky and S. Redner,
Phys. Rev. Lett.
 {\bf 90},
 238701
 (2003).

\bibitem{mob_red_03}
M. Mobilia, S. Redner,
Phys. Rev. E
 {\bf 68},
 046106
 (2003).

\bibitem{degroot_74}
M. H. DeGroot,
Journal of the American Statistical Association
 {\bf 69},
 118
 (1974).

\bibitem{berger_81}
R. L. Berger,
Journal of the American Statistical Association
 {\bf 76},
 415
 (1981).

\bibitem{axelrod_97}
R. Axelrod,
J. Conflict Resolution
 {\bf 41},
 203
 (1997).

\bibitem{cas_mar_ves_00a}
C. Castellano, M. Marsili, and A. Vespignani,
Phys. Rev. Lett.
 {\bf 85},
 3536
 (2000).

\bibitem{kle_egu_tor_mig_02}
K. Klemm, V. M. Egu\'{\i}luz, R. Toral, and M. San Miguel,
cond-mat/0210173.

\bibitem{vaz_kra_red_02}
F. Vazquez, P. L. Krapivsky, and S. Redner,
J. Phys. A: Math. Gen.
 {\bf 36},
 L61
 (2003).

\bibitem{kle_egu_tor_mig_03a}
K. Klemm, V. M. Egu\'\i luz, R. Toral, and M. San Miguel,
Physica A
 {\bf 327},
 1
 (2003).

\bibitem{kle_egu_tor_mig_03b}
K. Klemm, V. M. Egu\'{\i}luz, R. Toral, and M. San Miguel,
Phys. Rev. E
 {\bf 67},
 045101
 (2003).

\bibitem{vaz_red_04}
F. Vazquez and S. Redner,
J. Phys. A: Math. Gen.
 {\bf 37},
 8479
 (2004).

\bibitem{jacobmeier_05}
D. Jacobmeier,
Int. J. Mod. Phys. C
 {\bf 16},
 633
 (2005).

\bibitem{kle_egu_tor_mig_05}
K. Klemm V. M. Egu\'{\i}luz, R. Toral, and M. San Miguel,
Journal of Economic Dynamics and Control
 {\bf 29},
 321
 (2005).

\bibitem{gon_cos_tuc_05}
J. C. Gonz\'alez-Avella, M. G. Cosenza, and K. Tucci,
Phys. Rev. E
 {\bf 72},
 065102
 (2005).

\bibitem{kuperman_06}
M. N. Kuperman,
Phys. Rev. E
 {\bf 73},
 046139
 (2006).

\bibitem{gon_egu_cos_etal_06}
J. C. Gonz\'alez-Avella, V. M. Egu\'{\i}luz, M. G. Cosenza, K. Klemm, J. L. Herrera, and M. San Miguel,
Phys. Rev. E
 {\bf 73},
 046119
 (2006).

\bibitem{vaz_red_06}
F. Vazquez and S. Redner,
Europhys. Lett.
 {\bf 78},
 18002
 (2007).

\bibitem{def_nea_amb_wei_00}
G. Deffuant, D. Neau, F. Amblard, and G. Weisbuch,
Adv. Compl. Syst.
 {\bf 3},
 87
 (2000).

\bibitem{wei_def_amb_nad_01}
G. Weisbuch, G. Deffuant, F. Amblard, and J. P. Nadal,
cond-mat/0111494.

\bibitem{def_amb_wei_fau_02}
G. Deffuant, F. Amblard, G. Weisbuch, and T. Faure,
J. Artif. Soc. Soc. Simulation
 {\bf 5},
 4 $<$http://jasss.soc.surrey.ac.uk/5/4/1.html$>$
 (2002).

\bibitem{benn_kra_red_02}
E. Ben-Naim, P. L. Krapivsky, and S. Redner,
Physica D
 {\bf 183},
 190
 (2003).

\bibitem{stauffer_02}
D. Stauffer,
Int. J. Mod. Phys. C
 {\bf 13},
 315
 (2002).

\bibitem{sta_sou_sch_03}
D. Stauffer, A. O. Sousa, and C. Schulze,
cond-mat/0310243.

\bibitem{weisbuch_04}
G. Weisbuch,
Eur. Phys. J. B
 {\bf 38},
 339
 (2004).

\bibitem{sta_mey_04}
D. Stauffer and H. Meyer-Ortmanns,
J. Mod. Phys. C
 {\bf 15},
 241
 (2004).

\bibitem{def_amb_wei_04}
G. Deffuant, F. Amblard, and G. Weisbuch,
cond-mat/0410199.

\bibitem{amb_def_04}
F. Amblard, and G. Deffuant,
Physica A
 {\bf 343},
 725
 (2004).

\bibitem{assmann_04}
P. Assmann,
Int. J. Mod. Phys. C
 {\bf 15},
 1439
 (2004).

\bibitem{fortunato_04}
S. Fortunato,
Int. J. Mod. Phys. C
 {\bf 15},
 1301
 (2004).

\bibitem{fortunato_05}
S. Fortunato,
Int. J. Mod. Phys. C
 {\bf 16},
 17
 (2005).

\bibitem{wei_def_amb_05}
G. Weisbuch, G. Deffuant, and F. Amblard,
Physica A
 {\bf 353},
 555
 (2005).

\bibitem{lorenz_07}
J. Lorenz,
Int. J. Mod. Phys. C
 {\bf 18},
 1819
 (2007).

\bibitem{heg_kra_02}
R. Hegselmann and U. Krause,
J.~Artif.~Soc.~Soc.~Simulation
 {\bf 5},
 $<$http://jasss.soc.surrey.ac.uk/5/3/2.html$>$
 (2002).

\bibitem{plu_lat_rap_05}
A. Pluchino, V. Latora, and A. Rapisarda,
Eur. Phys. J. B
 {\bf 50},
 169
 (2006).

\bibitem{fortunato_04b}
S. Fortunato,
Int. J. Mod. Phys. C
 {\bf 15},
 1021
 (2004).

\bibitem{fortunato_04a}
S. Fortunato,
Physica A
 {\bf 348},
 683
 (2005).

\bibitem{fortunato_05a}
S. Fortunato,
Int. J. Mod. Phys. C
 {\bf 16},
 259
 (2005).

\bibitem{fortunato_05b}
S. Fortunato,
cond-mat/0501105.

\bibitem{for_sta_05}
S. Fortunato and D. Stauffer,
in: {\it Extreme Events in Nature and Society}, Eds. S. Albeverio, V. Jentsch, and H. Kantz,
 p. 233
 (Springer, Berlin, 2006).

\bibitem{for_lat_plu_rap_05}
S. Fortunato, V. Latora, A. Pluchino, and A. Rapisarda,
Int. Jour. Mod. Phys. C
 {\bf 16},
 1535
 (2005).

\bibitem{heg_kra_06}
R. Hegselmann and U. Krause,
J.~Artif.~Soc.~Soc.~Simulation
 {\bf 9},
 $<$http://jasss.soc.surrey.ac.uk/9/3/10.html$>$
 (2006).

\bibitem{lorenz_07a}
J. Lorenz,
arXiv:0708.3293
 (2007).

\bibitem{lorenz_08}
J. Lorenz,
Complexity
 {\bf 15},
 No. 4, p. 43
 (2010).

\bibitem{lorenz_08a}
J. Lorenz,
arXiv:0806.1587
 (2008).

\bibitem{lorenz_06}
J. Lorenz,
European Journal of Economic and Social Systems
 {\bf 19},
 213
 (2006).

\bibitem{lorenz_07b}
J. Lorenz,
thesis, Universit\"at Bremen
 (2007).

\bibitem{lorenz_05}
J. Lorenz,
Physica A
 {\bf 355},
 217
 (2005).

\bibitem{lorenz_07c}
J. Lorenz,
in: {\it Positive Systems, Lecture Notes in Control and Information Sciences, Vol. 341},
 209
 (Springer, Berlin, 2006).

\bibitem{lor_lor_08}
J. Lorenz and D. A. Lorenz,
IEEE Transactions on Automatic Control
 {\bf 55},
 1651
 (2010).

\bibitem{urb_lor_07}
D. Urbig and J. Lorenz,
in: {\it Proceedings of the Second Conference of the European Social Simulation Association (ISBN 84-688-7964-9)}
 (2004).

\bibitem{lag_abr_zan_04}
M. F. Laguna, G. Abramson, and D. H. Zanette,
Complexity
 {\bf 9},
 No. 4 p. 31
 (2004).

\bibitem{por_bol_sti_07}
M. Porfiri, E. M. Bollt, and D. J. Stilwell,
Eur. Phys. J. B
 {\bf 57},
 481
 (2007).

\bibitem{malarz_06}
K. Malarz,
Int. J. Mod. Phys. C
 {\bf 17},
 1521
 (2006).

\bibitem{lag_abr_zan_03}
M. F. Laguna, G. Abramson, and D. H. Zanette,
Physica A
 {\bf 329},
 459
 (2003).

\bibitem{pin_tor_her_09}
M. Pineda, R. Toral and E. Hern\'andez-Garc\'{\i}a,
J. Stat. Mech.
 P08001
 (2009).

%
%
\end{thebibliography}
\end{document}